\documentclass[letterpaper, 10 pt, conference]{ieeeconf}

\usepackage{cite}

\usepackage{amsthm}
\usepackage{amssymb}
\usepackage{amsfonts,amsmath}
\usepackage{mathtools}
\usepackage[font=footnotesize]{caption}
\usepackage{subcaption}
\usepackage{graphicx}
\usepackage{color, soul}
\usepackage{url}

\usepackage[usenames,x11names,table]{xcolor}
\usepackage{xspace} 
\usepackage{hyperref}
\usepackage{cleveref}
\usepackage{accents}
\usepackage{lipsum}

\usepackage[linesnumbered,vlined,ruled]{algorithm2e}
\usepackage{multirow} 
\usepackage{array} 

\newtheorem{problem}{Problem}

\newtheorem{remark}{Remark}
\newtheorem{definition}{Definition}
\newtheorem{example}{Example}
\newtheorem*{excont}{Example \continuation}
\newcommand{\continuation}{??}
\newenvironment{continueexample}[1]
 {\renewcommand{\continuation}{\ref{#1}}\excont[\textit{Cont'd}]}
 {\endexcont}

\title{\LARGE \bf Control Synthesis using Signal Temporal Logic Specifications with Integral and Derivative Predicates}

\author{Ali Tevfik Buyukkocak, Derya Aksaray, and Yasin Yaz{\i}c{\i}o\u{g}lu

\thanks{A.T. Buyukkocak and D. Aksaray are with the Department of Aerospace Engineering and Mechanics, University of Minnesota, Minneapolis, MN, 55455, {\tt\small buyuk012@umn.edu, daksaray@umn.edu}, and Y. Yaz{\i}c{\i}o\u{g}lu is with the Department of Electrical and Computer Engineering, University of Minnesota, Minneapolis, MN, 55455, {\tt\small ayasin@umn.edu}}
}

\IEEEoverridecommandlockouts
\begin{document}
\bibliographystyle{IEEEtran}
\maketitle
\thispagestyle{empty}
\pagestyle{empty}

\begin{abstract} \label{abstract}
In many applications, the integrals and derivatives of signals carry valuable information (e.g., cumulative success over a time window, the rate of change) regarding the behavior of the underlying system. In this paper, we extend the expressiveness of Signal Temporal Logic (STL) by introducing predicates that can define rich properties related to the integral and derivative of a signal. For control synthesis, the new predicates are encoded into mixed-integer linear inequalities and are used in the formulation of a mixed-integer linear program to find a trajectory that satisfies an STL specification. We discuss the benefits of using the new predicates and illustrate them in a case study showing the influence of the new predicates on the trajectories of an autonomous robot.  
\end{abstract}


\section{Introduction} \label{introduction}

\textcolor{black}{Motion planning and control of cyber-physical systems often require the satisfaction of complex tasks. One way of expressing such complex tasks is via temporal logics \cite{baier2008}. For example, Linear Temporal Logic (LTL) \cite{pnueli1977temporal} has been extensively used in planning and control of autonomous robots (e.g., \cite{kress2009temporal,karaman,guo2015multi,aksaray2015}).}

Temporal logics such as Metric Temporal Logic (MTL) \cite{koymans1990specifying} and Signal Temporal Logic (STL) \cite{maler},\cite{donze} are expressive specification languages that can define properties of dense-time real-valued signals with explicit spatial and time parameters. Different than the existing temporal logics with an automaton representation, MTL and STL contain \emph{predicates} in the form of inequalities and  are endowed with a metric called robustness degree that can quantify how good a signal satisfies a specification \cite{donze}. Robustness degree not only gives a yes/no answer but also provides a real value that quantifies the degree of satisfaction. 
Such a metric also enables to formulate an optimization problem that solves for a trajectory satisfying the temporal logic specifications (e.g., \cite{karaman,raman,pant2017smooth,buyukkocak2021distributed}).


Standard STL capabilities are enhanced by several studies in the literature (e.g., \cite{akazaki2015time,rodionova2016temporal,silvetti2018signal,xu2016census,sadigh2016safe}). For example, since conventional robustness degree focuses on critical time instants, authors of \cite{akazaki2015time} and \cite{rodionova2016temporal} define new measures to differentiate the satisfaction of predicates achieved at multiple time instants from the instantaneous satisfaction of them. 
With a similar motivation, a temporal operator is defined in \cite{silvetti2018signal} that explicitly specifies how long a predicate must be satisfied. However, existing approaches do not accommodate predicates expressing cumulative success or local behavior over a time interval.
To this end, this paper introduces integral and derivative predicates for STL. 
While the derivative predicate enables to define properties for the rate of change of the signal, the integral predicate allows the definition of properties such as average or cumulative progress at desired time intervals. 

This paper is closely related to \cite{brim2014stl} which defines predicates to compare signal values at different time instants to find local extrema of the signal. Also in \cite{haghighi2019control, mehdipour2019arithmetic,lindemann2019average}, authors propose the notions of cumulative and average robustness metrics calculated via the signal values at different time steps. With these methods, while the system can satisfy a conventional predicate as long and robust as possible by maximizing the new metrics, it does not result in the satisfaction of cumulative properties such as getting a specific amount of reward/value within a given time window.
To this end, we propose to define new integral and derivative predicates that can take into account the signal values at different time steps to define cumulative and relative specifications within desired time windows. By the proposed predicates, the local and global signal characteristics can be controlled extensively without losing the existing STL capabilities.

\textcolor{black}{This paper is organized as follows. We provide the notation and an overview of STL in Sec. \ref{preliminaries}. In Sec. \ref{problem statement}, we motivate our approach, introduce the new predicates, and state the optimization problem that solves for the trajectories satisfying the specifications defined with the proposed syntax. Then we present mixed-integer linear program (MILP) encodings of the new predicates in Sec. \ref{solution approach}. Simulation results of a case study with an autonomous robot are presented in Sec. \ref{case study}.}

\section{Preliminaries} \label{preliminaries}

\subsection{Notation} \label{notation}
In this paper, $\mathbb{R}_{\ge0}$ refers to the set of nonnegative real numbers, and $\mathbb{R}^n$ denotes the set of n-dimensional real-valued vectors. We represent $l_1$ and  $l_2$ norms by the operators of $|\cdot|$ and $\|\cdot\|$. Right and left time derivatives of a function are denoted as $d(\cdot)/dt
^+$ and $d(\cdot)/dt^-$, respectively.

\subsection{Signal Temporal Logic} \label{stl}
Rich time series can compactly be expressed by the Signal Temporal Logic \cite{maler}. In this paper, we use the following STL fragment:
\begin{equation}
\phi::= \mu \;|\; \neg\phi \;|\; \phi_1\wedge\phi_2\;|\; F_{[t_1,t_2]} \phi,
\label{eq:grammar}
\end{equation}
where $t_1,t_2\in\mathbb{R}_{\ge0}$ are time bounds with $t_2\geq t_1$; $F_{[t_1,t_2]}$, $\neg$, $\wedge$ are finally (i.e., eventually), negation, and conjunction (i.e., and) operators, respectively; $\phi$ is an STL formula, and $\mu$ is a predicate in the inequality form such as $\mu=g(\mathbf{x})\geq\, c$ with a constant $c\in\mathbb{R}$, a signal $\mathbf{x}:\mathbb{R}_{\ge0} \to\mathbb{R}^n$, and a function $g:\mathbb{R}^n \to \mathbb{R}$. Remaining useful operators are generated from the others as follows: $ G_{[t_1,t_2]} \phi= \neg F_{[t_1,t_2]} \neg \phi$ is globally (i.e., always) operator, and $\phi_1\vee\phi_2=\neg(\neg\phi_1\wedge\neg\phi_2)$ is disjunction (i.e., or) operator. We can also define an implication operator as $\phi_1\Rightarrow\phi_2=\neg\phi_1\vee\phi_2$.

Let $\mathbf{x}_t$ denote the value of $\mathbf{x}$ at time $t$ where $\mathbf{x}$ represents the run (or trajectory) of the system. 
Satisfaction of an STL formula by the part of the signal starting from $t$, i.e., $(\mathbf{x},t)$, is determined as follows:
\begin{equation}
\ 
\begin{split}
(\mathbf{x},t) &\vDash \mu \Longleftrightarrow g(\mathbf{x}_t)\geq\, c, \\
(\mathbf{x},t) &\vDash \neg\mu \Longleftrightarrow \neg\big((\mathbf{x},t)\vDash \mu\big),\\
(\mathbf{x},t) & \vDash \phi_1\wedge\phi_2 \Longleftrightarrow (\mathbf{x},t)\vDash \phi_1 \;\text{and}\; (\mathbf{x},t)\vDash \phi_2,\\
(\mathbf{x},t) & \vDash \phi_1\vee\phi_2 \Longleftrightarrow (\mathbf{x},t)\vDash \phi_1 \;\text{or}\; (\mathbf{x},t)\vDash \phi_2,\\
(\mathbf{x},t) & \vDash G_{[t_1,t_2]}\phi\Longleftrightarrow \forall\ t'\in[t+t_1,t+t_2],\;(\mathbf{x},t')\vDash \phi, \\
(\mathbf{x},t) & \vDash F_{[t_1,t_2]}\phi\Longleftrightarrow \exists\, t'\in[t+t_1,t+t_2],\;(\mathbf{x},t')\vDash \phi.
\end{split}
\label{eq:stl_semantics}
\end{equation}

While $(\mathbf{x},t)\vDash F_{[t_1,t_2]}\phi$ implies that $\phi$ must hold at least in one time instant between $[t+t_1,t+t_2]$, $(\mathbf{x},t)\vDash G_{[t_1,t_2]}\phi$ requires the satisfaction of $\phi$ at all time instants within the same interval. 
The horizon of an STL formula $\phi$, i.e., $hrz(\phi)$, can be defined as the minimum amount of time required to decide whether the formula is satisfied \cite{dokhanchi}. \color{black} Formally, $hrz(\phi)$ is found as:
\begin{equation}
\ 
\begin{split}
\mu &=g(\mathbf{x})\geq\, c \Longrightarrow hrz(\mu)=0, \\
\phi &=\neg \varphi \Longrightarrow hrz(\phi)=hrz(\varphi), \\
\phi &=\bigwedge_{i=1}^{m} \varphi _{i}\;\;\text{or}\;\;   \bigvee_{i=1}^{m} \varphi _{i} \Longrightarrow hrz(\phi) =\max_{i \in \{1,\dots, m\}} hrz (  \varphi _{i}),\\
\phi &=G_{[t_1,t_2]}\mu\;\;\text{or}\;\;\phi =F_{[t_1,t_2]} \mu   \Longrightarrow hrz(\phi) =t_2,\\
\phi &=G_{[t_1,t_2]}\varphi\;\;\text{or}\;\;\phi =F_{[t_1,t_2]} \varphi   \Longrightarrow hrz(\phi) =t_2 + hrz(\varphi).
\label{eq:horizon}
\end{split}
\end{equation}
\color{black}
For instance, the formula $G_{[0,5]}F_{[0,4]}x\geq0$ has a horizon of $5+4=9$, and the formula $G_{[0,5]} x\geq 0 \wedge F_{[0,4} x \geq 10$ has a horizon of $\max(5,4)=5$.

\normalsize
The satisfaction of specifications is indicated as either \textit{True} or \textit{False} for the most of the temporal logics. STL, on the other hand, is endowed with a metric called \emph{robustness degree}, $r(\mathbf{x}, \phi ,t) \in \mathbb{R}$, a real-valued function that is used to quantify the satisfaction of an STL formula $\phi$ with respect to a signal $(\mathbf{x},t)$. While positive robustness degree indicates the satisfaction of $\phi$, negative one represents a violation. In general, zero robustness degree is considered inconclusive, but we consider this case as satisfaction in this paper. The robustness degree metric can be formally and recursively defined as follows \cite{donze}:
\begin{equation}
\ 
\begin{split}
r(\mathbf{x},g(\mathbf{x})\geq \, c,t ) &=g(\mathbf{x}_t)-c, \\
r(\mathbf{x},\neg( g(\mathbf{x})\geq \, c) ,t) &=- r (\mathbf{x},g(\mathbf{x})\geq \, c,t), \\
r(\mathbf{x},\phi_{1}\wedge\phi_{2},t)& =\min\big(r(\mathbf{x},\phi_{1},t),r (\mathbf{x},\phi_{2},t) \big), \\
r(\mathbf{x},\phi_{1}\vee\phi_{2},t)& =\max \big(r(\mathbf{x},\phi_{1},t ) ,r  (\mathbf{x}, \phi _{2},t)\big), \\
r(\mathbf{x},F_{[t_1,t_2]}\phi,t)&=\mathop{\max }_{\mathop{t}^{'}\in[t+t_1,t+t_2]}r(\mathbf{x}, \phi ,t'),\\
r(\mathbf{x},G_{[t_1,t_2]}\phi,t)&=\mathop{\min }_{\mathop{t}^{'}\in[t+t_1,t+t_2]}r(\mathbf{x}, \phi ,t').
\end{split}
\label{eq:robustness_degree}
\end{equation}

\begin{figure}[htb!]
    \centering
    \includegraphics[trim=56 0 44 25,clip, width=.85\linewidth]{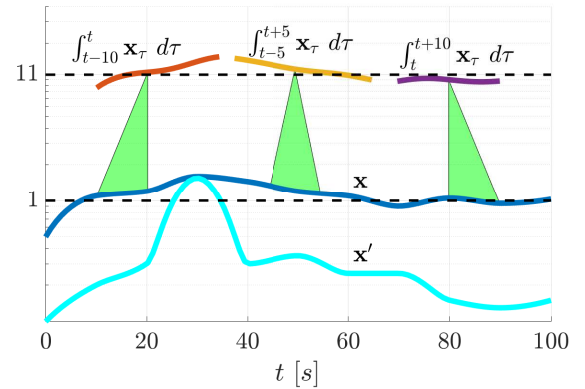}
    \caption{Signal functions $\mathbf{x}$ and $\mathbf{x}^\prime$ in blue and cyan, respectively in logarithmic scale, and portions of the curves generated by integrating $\mathbf{x}$ over three different relative time intervals. The values in the red, orange, and magenta curves are generated from the previous, closest, and next $10\ s$ of the blue $\mathbf{x}$ curve, respectively.}
    \label{fig:example}
    \vspace{-3mm}
\end{figure}

Robustness degree has limitations due to its main focus on the critical time instances and neglecting the remaining parts of the signal. For instance, suppose that eventually the value of $x$ needs to be at least $1$ within $[0,100]$, i.e., $\phi=F_{[0,100]}\ x\geq1$. Consider two trajectories $\mathbf{x}$ and $\mathbf{x}^\prime$ shown in Fig. \ref{fig:example}. Since, the degree of satisfaction is evaluated over any critical instant at which the predicate $\mathbf{x}_t\geq1$ is satisfied for $t\in[0,100]$, although the signal $\mathbf{x}$ satisfies the predicate for a longer time, robustness degree of both signals would be equal, i.e., $r(\mathbf{x},\phi,0)=r(\mathbf{x}^\prime,\phi,0)$. Therefore, the conventional robustness degree cannot differentiate between such cases and cannot provide comprehensive information about the signals $\mathbf{x}$ and $\mathbf{x}^\prime$. To address this issue, \cite{akazaki2015time, rodionova2016temporal,silvetti2018signal} define new measures that can capture the duration of predicate satisfaction. While existing metrics can track how long a predicate $x(t)\geq \delta$ is satisfied within an interval $[0,T]$, they are not able to address a notion of cumulative success such as $\int_0^t x(\tau) d\tau \geq \delta$ within an interval $[0,T]$.
\section{Problem Statement} \label{problem statement}
\subsection{Motivation} \label{motivation}
While the duration of satisfaction of a conventional predicate may be important for certain applications; it may also be desired to meet some cumulative success criteria. For instance, in \cite{xu2016census} cumulative properties are defined over a swarm. Authors in \cite{haghighi2019control,mehdipour2019arithmetic,lindemann2019average}, propose discrete-time cumulative and average robustness degrees calculated by the summation of the robustness degrees of the same predicate at different time steps. This enables to have a predicate satisfied as long and robust as possible. 
In \cite{brim2014stl}, a freezing operator is used to store the signal values at different time instants in memory and to compare them inside the predicates which enables to find local extrema and examine the oscillatory behavior. In other words, the satisfaction of a predicate now depends on more than one signal values at different time instants. Similarly, in the case of progressive events, one may desire to evaluate the contributions of the past, future, or both signal values to determine a satisfaction at the current time. Therefore, the success may depend on the accumulation of signal values from different time instants. We can define such success criteria inside the same predicate, and use the integral of a signal over a given bounded time interval to assess satisfaction.


\begin{example}
Consider a continuous signal function that represents the speed, $\|v\|$. Let trajectory of $\|v\|$ be the same with the signal $\mathbf{x}$ in Fig. \ref{fig:example}. Assume that at $t=20$, system is desired to travel more than $11\ m$ within the last $10\ s$. In other words, integration of $\|v\|$ over $[10,20]$ must be at least $11$. One may check the satisfaction of this specification at $t=20$ by inspecting the red curve in Fig. \ref{fig:example} obtained by the integration of $\|v\|$ (i.e., $\mathbf{x}$) over the window of the previous $10\ s$ at each time instant.

The interval of interest over which the integrated signal may not depend only on the past values. One may also specify thresholds on the integrals defined partially or fully over the future signal values. In Fig. \ref{fig:example}, sample intervals defined in the past, future, or both are partially shown with red, magenta, and orange, respectively. For each time instant, the window of $10\ s$ is shifted throughout the time axis to generate the integral curves. While at $t=50$ total distance traveled between $[45,55]$ (depending on both past and future values) is more than $11\ m$, at $t=80$ this time the distance traveled during $[80,90]$ (depending on only future values) is less than the given threshold. 
\label{example_signal}
\end{example}

In the next section, we introduce STL predicates whose satisfaction can depend on part of the past, present, or future of the signal. The new predicates are not the same as using the standard predicates with temporal operators. For example, consider some specifications defined between $t=80$ and $t=90$ over $\mathbf{x}$ represented in Fig. \ref{fig:example}. Trajectory of $(\mathbf{x},80)$ does not satisfy the predicate $x\geq5$ because $\mathbf{x}_{80}<5$. Moreover, the trajectory $(\mathbf{x},0)$ does not satisfy an STL formula $F_{[80,90]}x\geq5$ because there is no $t\in[80,90]$ such that $\mathbf{x}_t\geq 5$. Now, consider another predicate including an integral over time whose satisfaction does not depend only on the current time step but determined by the signal values within the bounds of the integral. The same signal $(\mathbf{x},0)$ satisfies such a predicate $\int_{80}^{90} \mathbf{x}_\tau d\tau\geq 5$ as shown in Fig. \ref{fig:example} with magenta. Furthermore, temporal operators can also be used with such a predicate to create more complex specifications by shifting the integral interval.

One may try to represent such specifications by simply considering the integral or derivative of the original signal, and using it with conventional predicates. For example, again consider the magenta signal in Fig. \ref{fig:example}. Let the specification be ``the integrations of $\mathbf{x}$ within the time windows of $[t+a,t+b]$ need always to be at most $11$ for $t\in[70,90]$, i.e., $G_{[70,90]}\int_a^{b}\mathbf{x}_\tau d\tau\leq11$. This specification could be expressed by first defining a new signal depicting the integral of $\mathbf{x}$, i.e., $H(t)=\int_0^{t} \mathbf{x}_\tau d\tau$. Then, we can write $G_{[70,90]}H(t+b) - H(t+a)\leq11$. 
While such predicates with multiple time instants  (i.e., $H(t+b) - H(t+a)$) are not commonly used, they can be accommodated by the syntax and semantics of STL \cite{donze2012temporal}.
Alternatively, we 
introduce an \textit{integral predicate} that can explicitly define certain progress over the given time interval, together with a \textit{derivative predicate} to evaluate local characteristics of the signal. These predicates take time bounds as input and shift these bounds in accordance with the outer temporal operators. Accordingly, they do not require creating the integral and derivative of $\mathbf{x}$ as additional signals. 

\color{black}

\subsection{Definition of Integral and Derivative Predicates} \label{new_temporal_predicates}

We first introduce the integral predicate to express a specific amount of progress in preemptable and cumulative properties via STL specifications.

\begin{definition}
(Integral predicate) An integral predicate over a bounded time interval $[a,b]\subset\mathbb{R}$ with $b>a$ is defined as:
\begin{equation}
\ 
\mu^i_{[a,b]} = \int_{a}^{b}g(\mathbf{x}_\tau)d\tau \geq c,
\label{eq:temporal_pred_def}
\end{equation}
\noindent where $c\in\mathbb{R}$ is a constant, $\mathbf{x}:\mathbb{R}_{\ge0} \to\mathbb{R}^n$ is the signal, and $g:\mathbb{R}^n \to \mathbb{R}$ is an integrable function.
\label{definition_temp_pred}
\end{definition}

Notice that the integral predicate requires the input of the time bounds as the temporal operators like eventually and globally. With the new integral predicate, one may explicitly define these integral bounds, and specify some progress threshold over a certain time interval. 

\begin{continueexample}{example_signal}
We can represent the specifications mentioned in Example \ref{example_signal} and shown in Fig. \ref{fig:example} with the integral predicates as follows. Both $\mu^i_{[-10,0]}(t)=\int_{t-10}^{t}\|v\|\ d\tau \geq11$ defined on the past values (in red) and $\mu^i_{[-5,5]}(t)=\int_{t-5}^{t+5}\|v\|\ d\tau \geq11$ depending on both past and future values (in orange) are satisfied for $t=20$ and $t=50$, respectively.  However, for $t=80$ the integral predicate $\mu^i_{[0,10]}(t)=\int_{t}^{t+10}\|v\|\ d\tau \geq11$ that uses a future time interval (in magenta) is violated.
\end{continueexample} 

In addition to cumulative properties, it is possible to have volatility over the signal. To explicitly bound such behavior, we introduce the derivative predicate.
\begin{definition}
(Derivative Predicate) A derivative predicate is defined to specify the queries on the rate of change of the signal function $g(\mathbf{x})$ with the time by the first derivative of it as follows: 
\begin{equation}
\ 
    \mu^d_+ = \frac{dg(\mathbf{x})}{dt^+}\geq c \quad\quad\bigg| \quad\quad \mu^d_- = \frac{dg(\mathbf{x})}{dt^-}\geq c,
\end{equation}
\noindent
where $\mu^d_+$ and $\mu^d_-$ denote the right and left derivative.
\label{definition_derivative_pred}
\end{definition}

By using the newly defined predicates together with the standard STL predicates, we can define diverse specifications related to the local and global characteristics of the signals accordingly to the following STL syntax:
\begin{equation}
\phi::= \mu \;|\; \mu^i_{[a,b]} \;|\; \mu^d_{^{+\!,-}} \;|\; \neg\phi \;|\; \phi_1\wedge\phi_2\;|\; F_{[a,b]} \phi,
\label{eq:temporal_grammar}
\end{equation}

\noindent where the satisfactions of $\mu^i_{[a,b]}$ and $\mu^d_{^{+\!,-}}$ are determined for any signal $\mathbf{x}:\mathbb{R}_{\geq0}\to\mathbb{R}^n$ as follows:
\begin{equation}
\ 
\begin{split}
(\mathbf{x},t) &\vDash \mu^i_{[a,b]} \Longleftrightarrow \int_{t+a}^{t+b}g(\mathbf{x}_\tau)d\tau \geq c, \\
(\mathbf{x},t) &\vDash \mu^d_{^{+}} \Longleftrightarrow \frac{dg(\mathbf{x}_t)}{dt^{^{+}}} \geq c, \\
(\mathbf{x},t) &\vDash \mu^d_{^{-}} \Longleftrightarrow \frac{dg(\mathbf{x}_t)}{dt^{^{-}}} \geq c, \\
(\mathbf{x},t) &\vDash \neg\mu^i_{[a,b]} \Longleftrightarrow \neg\big((\mathbf{x},t)\vDash \mu^i_{[a,b]}\big),\\
(\mathbf{x},t) &\vDash \neg\mu^d_{^{+\!,-}} \Longleftrightarrow \neg\big((\mathbf{x},t)\vDash \mu^d_{^{+\!,-}}\big).
\end{split}
\label{eq:semantics}
\end{equation}

Note that the signal $\mathbf{x}$ is undefined for $t<0$, therefore we assume $t+a\geq 0$ throughout the paper. By preserving the quantitative semantics for the common operators defined in \eqref{eq:robustness_degree}, we can quantify the satisfaction of the new predicates similarly as:  
\begin{equation}
\ 
\begin{split}
r(\mathbf{x},\mu^i_{[a,b]},t ) &=\int_{t+a}^{t+b}g(\mathbf{x}_\tau)d\tau- c, \\
r(\mathbf{x},\mu^d_{^{+}},t ) &=\frac{dg(\mathbf{x}_t)}{dt^{^{+}}} - c, \\
r(\mathbf{x},\mu^d_{^{-}},t ) &=\frac{dg(\mathbf{x}_t)}{dt^{^{-}}} - c, \\
r(\mathbf{x},\neg\mu^i_{[a,b]} ,t) &=- r (\mathbf{x}_t,\mu^i_{[a,b]},t),\\
r(\mathbf{x},\neg\mu^d_{^{+\!,-}} ,t) &=- r (\mathbf{x}_t,\mu^d_{^{+\!,-}},t).
\end{split}
\label{eq:temporal_robustness_degree}
\end{equation}

\color{black}
Again a nonegative robustness degree indicates the satisfaction of the predicates, e.g., $r(\mathbf{x},\mu^i_{[a,b]} ,t) \geq 0 \Rightarrow  (\mathbf{x},t ) \vDash \mu^i_{[a,b]}$, while negative one represents a violation ($r(\mathbf{x}, \mu^i_{[a,b]} ,t) <0 \Rightarrow  (\mathbf{x},t ) \nvDash \mu^i_{[a,b]})$.


\begin{remark}
In STL control synthesis, the satisfaction of temporal operators such as \emph{eventually} and \emph{globally} at $t$ are generally decided  by the assessment of the future time steps $t'\geq t$. An integral predicate $\mu^i_{[a,b]}$ for $[a,b]\subset\mathbb{R}$, or the derivative one, $\mu^d_{^{+,-}}$, can depend both on the past and future signal values by enabling negative time bounds.
\end{remark}

\color{black}
An integral predicate can have negative time bounds with respect to the bounds of the outer temporal operators (e.g., eventually and globally). We can also use the integral predicate alone by using nonnegative time bounds or defining it with the negative time bounds at future time steps. The horizon definition of STL \cite{dokhanchi} can be extended for the integral predicate and nested use of it with the temporal operators as follows:
\color{black}
\begin{equation}
\ 
\begin{split}
\mu^i_{[a,b]} &=\int_{a}^{b}g(\mathbf{x}_\tau)d\tau\geq c \\
&\quad\quad\quad\Longrightarrow hrz(\mu^i_{[a,b]})=\max\big(|a|,b,b-a\big), \\
\phi &=G_{[t_1,t_2]}\mu^i_{[a,b]}\;\;\text{or}\;\;\phi =F_{[t_1,t_2]} \mu^i_{[a,b]} \\ &\quad\quad\quad\Longrightarrow hrz(\phi)=\max(t_2,t_2+b), 
\label{eq:temporal_horizon}
\end{split}
\end{equation}

\noindent where $t_1,t_2\in\mathbb{R}_{\geq0}$ and $a,b\in\mathbb{R}$ are time bounds with $t_2\geq t_1$, $b>a$, and a constraint of $t_1+a\geq0$ for the sake of nonnegative global time. 

\begin{remark}
For discrete-time signals, we define the integral and derivative predicates at time $t$ assuming a discrete-time signal function $g(\mathbf{x})$ as:
\begin{equation}
\ 
\begin{split}
&\mu^i_{[a,b]} = \sum_{k'=k+a/\delta t}^{k+b/\delta t-1}g(\mathbf{x}_{k'\delta t})\,\delta t \geq c\, ,\\
&\mu^d_+ \!= g(\mathbf{x}_{(k+1)\delta t})\!-\!g(\mathbf{x}_{k\delta t})\geq c\, \delta t\\
&\mu^d_- \!= g(\mathbf{x}_{k\delta t})\!-\!g(\mathbf{x}_{(k-1)\delta t})\geq c\, \delta t,
\end{split}
\label{eq:temporal_pred_def_disc}
\end{equation}

\noindent where $k\in \mathbb{Z}$ is the step number, and $\delta t$ is the time step such that $t=k\delta_t$. 

\end{remark}

Modified robustness metrics \cite{haghighi2019control,mehdipour2019arithmetic,lindemann2019average} can also be used to quantify cumulative properties. However, instead of modifying the semantics of existing temporal operators, we introduce the integral predicate as a new operator with its own qualitative and quantitative semantics that can easily be used with the standard STL syntax and semantics. 
For example, consider a signal $\mathbf{x}=\{1,1,1,1,1,2,\epsilon\}$, where $\epsilon$ is a small arbitrary number since $\mu^i_{[a,b]}$ does not consider the last signal value in discrete-time signals. Suppose that the sum of any two consecutive signal values need to eventually be at least $3$. The proposed integral predicate enables to define this as $\varphi=F_{[0,4]}\mu^i_{[0,2]}\geq3$ where $\mu^i_{[0,2]}=\int_0^2x\, dt$, for which the robustness degree of $x$ with respect to $\varphi$ is $r(\mathbf{x},\varphi)=0$. Thus, $x$ barely satisfies the specification $\varphi$. Now, consider the \emph{cumulative robustness degree}, $\rho^+(\cdot)$ \cite{haghighi2019control} that is the closest measure to our proposed idea. One can express the task in $\varphi$ by using the cumulative robustness degree as $\varphi'=F_{[0,4]}\big(\rho^+(\mathbf{x},F_{[0,1]}x\geq0)\geq3\big)$ for the same signal, which has $\rho^+(\mathbf{x},\varphi')=-4$ implying a violation while $\mathbf{x}$ actually satisfies $\varphi$. 

\color{black}
Overall, there is no specific syntax in STL that can capture the cumulative properties, and the proposed integral predicate is introduced to facilitate the definition of cumulative signal properties via STL.

\color{black}


\subsection{Optimal Control Problem}
\label{optimal_control_problem}

We consider a discrete time system $\mathrm{x}^+=f(\mathrm{x},\mathrm{u})$ evolving over a continuous state space with state and input vectors of $\mathrm{x}\in\mathbb{R}^n$ and $\mathrm{u}\in\mathbb{R}^m$, respectively. Accordingly, the signal we evaluate becomes the finite state trajectory, $\mathbf{x} = \big[\mathrm{x}_{0\delta t}\;\mathrm{x}_{1\delta t}\,\cdots\, \mathrm{x}_{H\delta t} \big]\in\mathbb{R}^{n\times (H+1)}$ where $H$ is the length of the mission horizon. 

The state trajectory, $\mathbf{x}$, is desired to achieve an STL specification, $\Phi$. Note that the mission horizon has to be longer than the specification horizon, i.e., $H \geq hrz(\Phi)/\delta t$ to determine the satisfaction or violation of $\Phi$.

\begin{problem} Given a system with discrete-time dynamics, $\mathrm{x}^+=f(\mathrm{x},\mathrm{u})$, find the optimal control policy over the horizon $\mathbf{u}^* = \big[\mathrm{u}^{\ast}_{0\delta t}\,\mathrm{u}^{\ast}_{1\delta t}\,\cdots\, \mathrm{u}^{\ast}_{(H-1)\delta t} \big]\in\mathbb{R}^{m\times H}$ to achieve a global specification $\Phi$:  
\begin{equation}
\ 
\begin{split}
\bold{u}^*=&arg\,min \;\sum _{k=0}^{H-1}\;\mathcal{J} \big(\mathrm{x}_{k\delta t},\mathrm{u}_{k\delta t}\big) \\
s.t.\;\;\;\;&\mathrm{x}^+=f(\mathrm{x},\mathrm{u}),\;\\&\mathbf{x}_0=\mathrm{x}_{0}, \; r( \mathbf{x}, \Phi ,0)  \geq 0
\end{split}
\label{eq:r_optimization}
\end{equation}

\noindent where $\mathcal{J} \big( \mathrm{x}_{k\delta t},\mathrm{u}_{k\delta t} \big)$ is a running cost as a function of state vector and control inputs, $\mathcal{J}:\mathbb{R}^n\times \mathbb{R}^m\to\mathbb{R}$; $\mathrm{x}_0$ is the initial state vector, and $r( \mathbf{x}, \Phi ,0)  \geq 0$ enforces the satisfaction of the temporal logic constraints by requiring the robustness degree to be nonnegative.
\label{problem:optimization}
\end{problem}

\section{Solution Approach} \label{solution approach}
The solution of \eqref{eq:r_optimization} requires a nonnegative robustness degree (\eqref{eq:robustness_degree} and \eqref{eq:temporal_robustness_degree}) to enforce satisfaction. However, it is computed via recursive definitions of computationally expensive and non-smooth $\min$ and $\max$ functions. Therefore, using it as a constraint in the optimization or feasibility problems makes the problem non-trivial. Alternatively, satisfaction of $\Phi$ can be encoded as a set of constraints with binary variables in the form of $z^{\Phi} [k]  \in  \{ 0,1 \}$ \cite{karaman},\cite{raman}. For each predicate in the form of inequality, a couple of big M constraints can be written depending on a binary variable as
\color{black}

\begin{equation}
\ 
\begin{matrix}
\mu =g(\mathbf{x}) \ge c\;\;\; \left\{\; \begin{matrix*}[l]
g(\mathbf{x})-c\ge M( z^{\mu}-1), \\
g(\mathbf{x})-c\le M z^{\mu}, \end{matrix*}\right.
\end{matrix}
\label{eq:big_M}    
\end{equation}

\noindent where $M \in \mathbb{R}^{+}$  is a sufficiently large number. Similarly, the satisfaction of an integral predicate can also be encoded as
\begin{equation}
\ 
\begin{matrix}
\phi =\mu^i_{[a,b]} \;\;\; \left\{\; \begin{split}
\sum_{k'=k+a/\delta t}^{k+b/\delta t-1}g(\mathbf{x}_{k'\delta t})\delta t-c\ge M( z^{\phi}[k]-1), \\
\sum_{k'=k+a/\delta t}^{k+b/\delta t-1}g(\mathbf{x}_{k'\delta t})\delta t-c\le M z^{\phi}[k]. \end{split}\right.
\end{matrix}
\label{eq:temporal_big_M}    
\end{equation} 

The derivative predicate is encoded in a similar fashion as follows:
\ 
$$
\mu^d_+ =\normalsize{\frac{dg(\mathbf{x})}{dt^+}}\small \ge c\;\left\{\begin{matrix*}[l]
g(\mathbf{x}_{(k+1)\delta t})-g(\mathbf{x}_{k\delta t})-c\,\delta t\\
\hspace{31mm} \ge M( z^{\mu^d_{^+}}[k]-1), \\
g(\mathbf{x}_{(k+1)\delta t})-g(\mathbf{x}_{k\delta t})-c\,\delta t\le M z^{\mu^d_{^+}}[k], \end{matrix*}\right.\\
$$
\begin{equation}
\ 
\mu^d_- =\frac{dg(\mathbf{x})}{dt^-} \ge c\;\left\{\begin{matrix*}[l]
g(\mathbf{x}_{k\delta t})-g(\mathbf{x}_{(k-1)\delta t})-c\,\delta t\\
\hspace{31mm}\ge M( z^{\mu^d_{^-}}[k]-1), \\
g(\mathbf{x}_{k\delta t})-g(\mathbf{x}_{(k-1)\delta t})-c\,\delta t\le M z^{\mu^d_{^-}}[k]. \end{matrix*}\right.
\label{eq:derivative_big_M}    
\end{equation}

\normalsize
Starting with the predicates, remaining binary constraints corresponding to an STL formula can be built into each other. \color{black}The connections of the Boolean operators with other temporal operators and predicates can be constructed with the following rules and encoded as integer constraints considering two sample formulas $\phi$ and $\varphi$ \cite{raman}.

\vspace{5mm}
\noindent
\textbf{\textit{Negation:}} $\phi=\neg \varphi$
\vspace{-2mm}
\begin{equation}
z^{ \phi}_k =1-z^{  \varphi }_k.
\label{eq:z_negation}
\end{equation}
\textbf{\textit{Conjunction:}} $\phi=\bigwedge_{i=1}^{m} \varphi _{i}$
\begin{equation}
\begin{split}
z^{\phi}_k&\leq z^{\varphi_{i}}_k,\;\;i=1, \ldots , m, \\
z^{\phi}_k&\geq 1-m+ \sum_{i=1}^{m}z^{\varphi _{i}}_k.
\end{split}
\label{eq:z_conjunction}
\end{equation}
\textbf{\textit{Disjunction:}} $\phi = \bigvee_{i=1}^{m} \varphi _{i}$
\begin{equation}
\begin{split}
z^{\phi}_k&\geq z^{\varphi_{i}}[k],\;\;i=1, \ldots , m, \\
z^{\phi}_k&\leq \sum_{i=1}^{m}z^{\varphi _{i}}_k.
\end{split}
\label{eq:z_disjunction}
\end{equation}\\

\noindent\textbf{\textit{Globally:}} $\phi =G_{[t_1,t_2]}\varphi$
\vspace{-4mm}
\begin{equation}
z^{ \phi}_k =\bigwedge_{ k'=k+t_1/\delta t}^{k+t_2/\delta t}z^{  \varphi }_{k'}.
\label{eq:z_globally}
\end{equation}
\textbf{\textit{Eventually:}} $\phi =F_{[t_1,t_2]} \varphi$
\vspace{-3mm}
\begin{equation}
\quad z^{ \phi}_k =\bigvee_{ k'=k+t_1/\delta t}^{k+t_2/\delta t}z^{  \varphi }_{k'}.
\label{eq:eventually}
\end{equation}

\color{black}
\normalsize 
As a result, the satisfaction of the general formula, $\Phi$, which is previously implied by $r(\mathbf{x},\Phi,k) \geq 0$, is now rendered to $z^{\Phi}_k =1$. 
Since we handle the specifications on the trajectory of the state vector, $\mathrm{x}_k$, inside the temporal logic constraints, in the optimization, we penalize only control inputs, $\mathrm{u}_k$. Therefore, we define the running cost as the $l_{1}-$norm of the control input, i.e., $\mathcal{J} \big( \mathrm{u}_k \big) = \big| \mathrm{u}_k  \big|$. 

Nonlinear systems can be linearized around known equilibrium points. Furthermore, the mission constraints can be encoded as linear inequalities with binary variables as in \eqref{eq:big_M}, \eqref{eq:temporal_big_M}, \eqref{eq:derivative_big_M}. Accordingly, under linear dynamics and predicates, the constrained optimization problem in \eqref{eq:r_optimization} can be posed as a mixed-integer linear program.
\begin{problem}
\begin{equation}
\ 
\begin{split}
\bold{u}^*=&arg\,min \;\sum _{k=0}^{H-1}\;\big| \mathrm{u}_k  \big| \\
s.t.\;\;\;\;&\mathrm{x}_{k+1}=A\mathrm{x}_k+B\mathrm{u}_k,\;\\&\mathbf{x}_0=\mathrm{x}_{0}, \; z^\Phi_0  = 1,
\end{split}
\label{eq:z_optimization}
\end{equation}
\noindent where $A\in\mathbb{R}^{n\times n}$ and $B\in\mathbb{R}^{n\times m}$ are the relevant system and input matrices, respectively.
\label{milp_problem}
\end{problem}

Note that solving Problem 2 enforces the satisfaction of the STL specification (due to the existence of constraint $z^\Phi_0  = 1$). It can be solved via several off-the-shelf tools such as MATLAB's built-in integer programming solver, \textit{intlinprog}, and Gurobi \cite{gurobi}. This problem can also be formulated for maximal satisfaction of the STL specification by relaxing each predicate with a slack variable to be minimized with the negation-free STL specifications \cite{akazaki2015time}, \cite{rodionova2016temporal}.

\section{Case Study} \label{case study}
\begin{figure*}[t]
     \centering
     \begin{subfigure}[b]{0.329\textwidth}
         \centering
         \includegraphics[trim=25 0 30 20,clip,width=\textwidth]{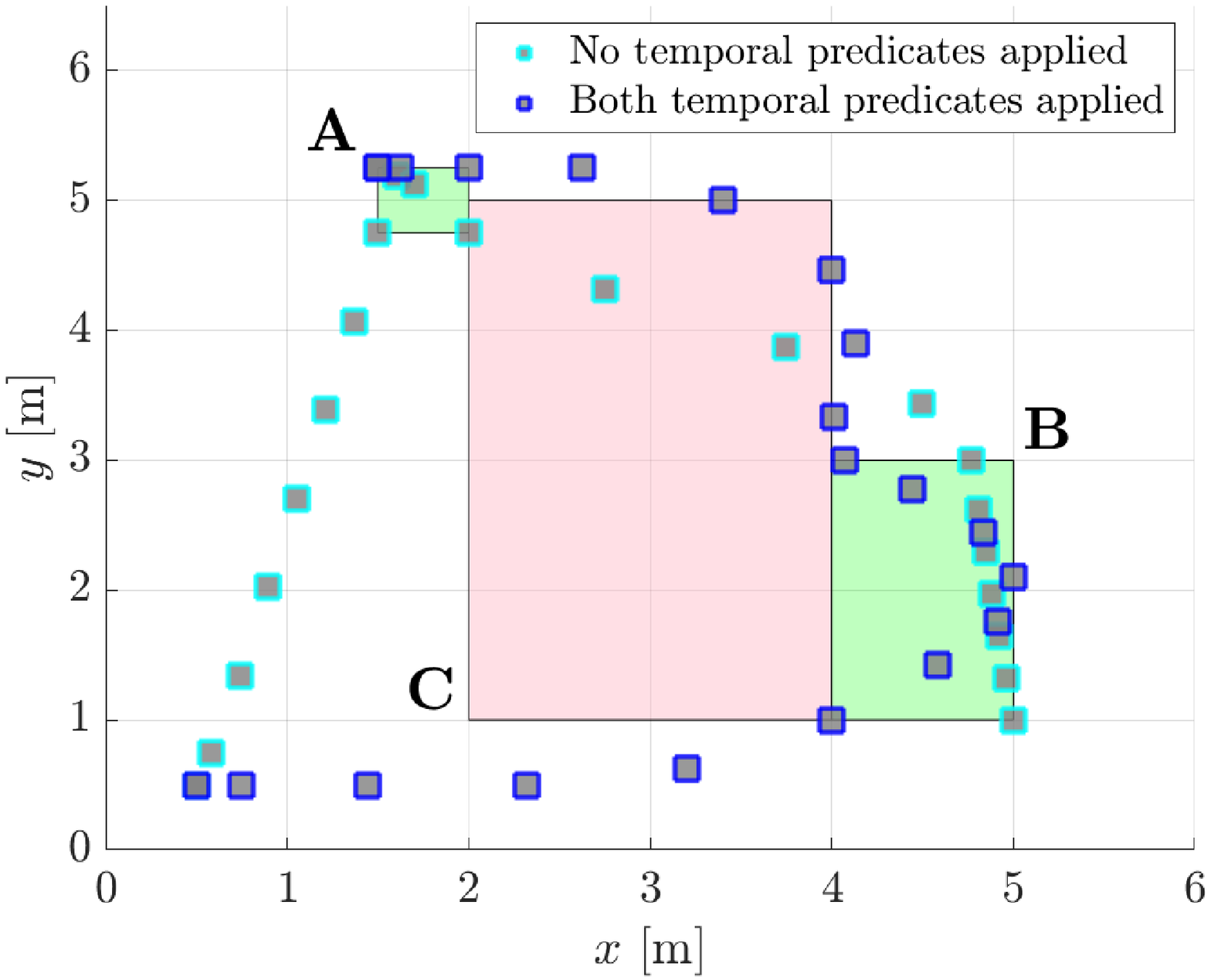}
         \caption{}
         \label{fig:sim_none}
     \end{subfigure}
     \begin{subfigure}[b]{0.329\textwidth}
         \centering
         \includegraphics[trim=25 0 30 20,clip,width=\textwidth]{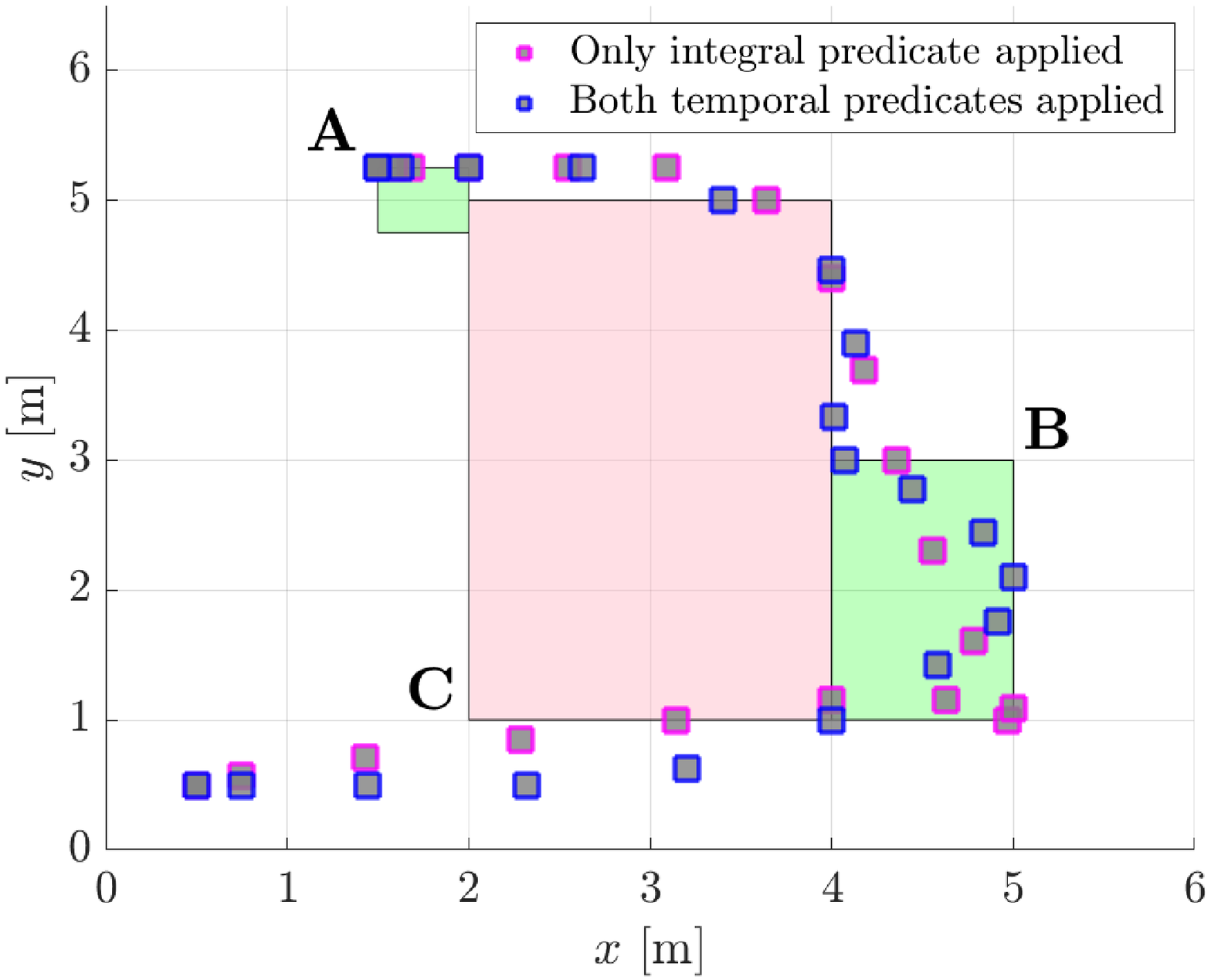}
         \caption{}
         \label{fig:sim_int}
     \end{subfigure}
     \begin{subfigure}[b]{0.329\textwidth}
         \centering
         \includegraphics[trim=25 0 30 20,clip,width=\textwidth]{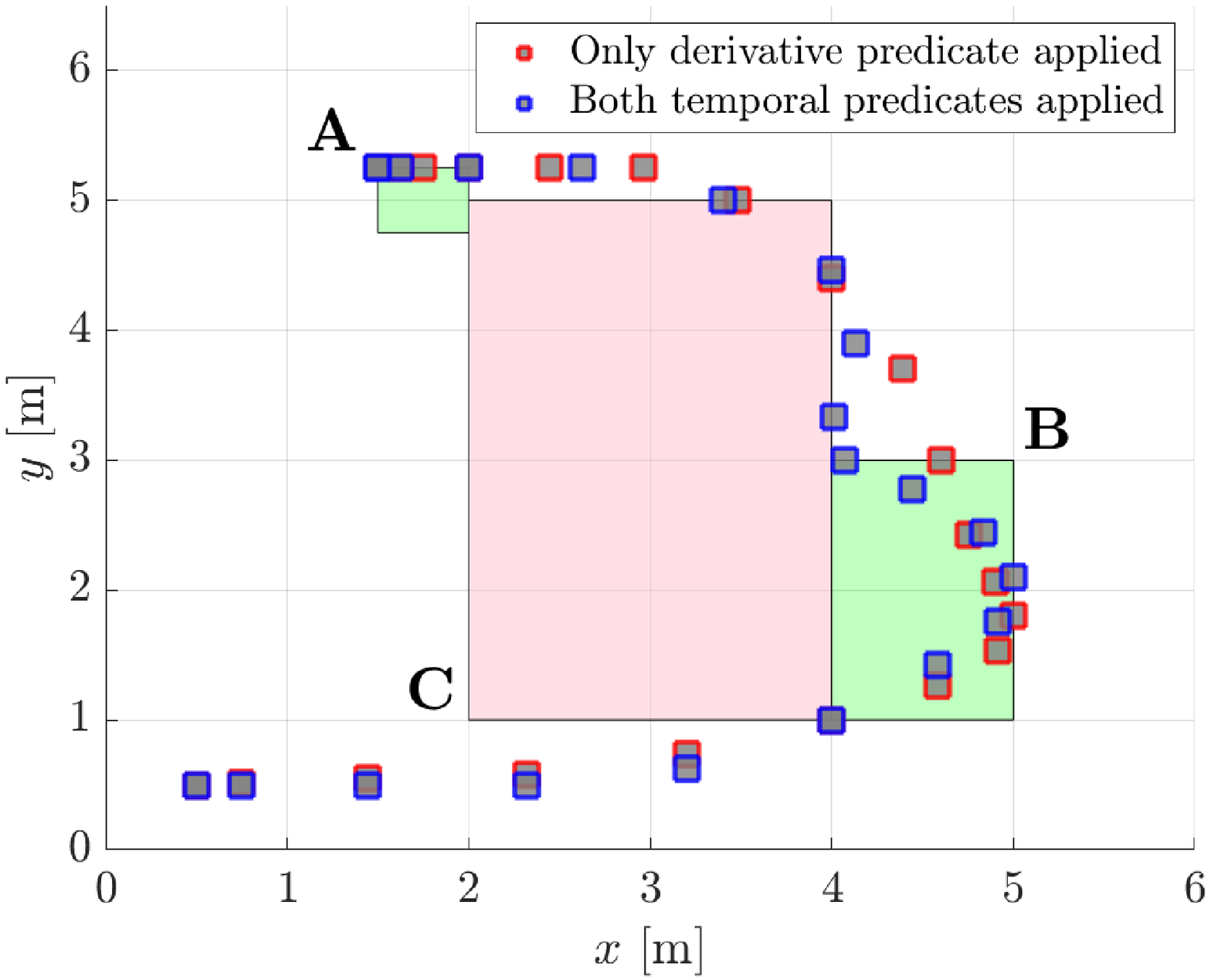}
         \caption{}
         \label{fig:sim_deriv}
     \end{subfigure}
        \caption{Comparisons of the blue trajectory that achieves the original specification \eqref{eq:formula} with the ones achieving the same specification except a) the integral and derivative predicates (cyan) b) the derivative predicates (magenta) c) the integral predicates (red). All trajectories start from $(0.5,0.5)\ m$.  
       }
        \label{fig:simulation}
        \vspace{-2mm}
\end{figure*}
To illustrate the the functionality of the new predicates, here we define a complementary example.
\begin{example}
Consider an autonomous robot with discrete-time double integrator dynamics 
\begin{equation}
\ 
\mathrm{x}_{k+1}=
   \begin{bmatrix}
      1 & \delta_t & 0 & 0 \\
      0 & 1 & 0 & 0 \\
      0 & 0 & 1 & \delta_t \\
      0 & 0 & 0 & 1
   \end{bmatrix} 
   \mathrm{x}_k +
   \begin{bmatrix}
   0.5 \delta_t^2 & 0\\
   \delta_t & 0\\
   0 & 0.5 \delta_t^2\\
   0 & \delta_t
   \end{bmatrix} \mathrm{u}_k,
   \label{eq:cont_dyn}
\end{equation}
with the state vector $\mathrm{x}=[x,v_x,y,v_y]^T$ where $v_x$, $v_y\in\mathbb{R}$ are the velocities in $x$, $y\in\mathbb{R}$ directions, respectively; and the input vector $\mathrm{u}=[u_x,u_y]^T$ where $u_x$,$u_y\in\mathbb{R}$ are the specific forces in given directions. The goal of the mission is monitoring the natural habitat in predefined areas by observing as much as possible without disturbing the members of it. Mission specifications are given as follows: i) Continuously service regions A and B for at least $3\ s$ and $6\ s$, respectively, and while in B, travel inside it for at least $2\ m$ in both directions. ii) Whenever in A or B, do not disturb the members of the natural habitat by limiting the acceleration to $0.25\ m/s^2$ in each direction. iii) Overall the robot cannot exceed an acceleration of $0.5\ m/s^2$ in any direction. iv) Since the habitants of region C are adversarious, whenever you are in C keep the horizontal velocity at least $1\ m/s$ to leave the C immediately (its shortest edge lies in the horizontal direction). With a total mission time of $20\ s$, we can express these specifications as follows:
\begin{equation}
\ 
    \begin{split}
    \Phi =\ &F_{[0,17]}\big(G_{[0,3]}R_A\big)\wedge F_{[0,14]}\big(G_{[0,6]}R_B\wedge\mu^{i,1}_{[0,6]}\wedge\mu^{i,2}_{[0,6]}\big)\\
    &\wedge G_{[1,20]}\big( (R_A\vee R_B) \Rightarrow (\mu^{d,1}_-\wedge\mu^{d,2}_-)\big)\\
    &\wedge G_{[0,19]}\big(\mu^{d,3}_+\wedge\mu^{d,4}_+\big)\\
    &\wedge G_{[0,20]}\big(R_C \Rightarrow |v_x|\geq1\big),
    \end{split}
    \label{eq:formula}
\end{equation}

\noindent where the integral predicates are defined to constrain the total traveled distance as $\mu^{i,1}_{ [0,6]}=\int^6_0 |v_x| d\tau\geq2$ and $\mu^{i,2}_{ [0,6]}=\int^6_0 |v_y| d\tau\geq2$; the derivative predicates limit the acceleration either temporarily or throughout the mission as $\mu^{d,1}_-=|\dot{v}_x|\leq0.25$, $\mu^{d,2}_-=|\dot{v}_y|\leq0.25$, $\mu^{d,3}_+=|\dot{v}_x|\leq0.5$, and $\mu^{d,4}_+=|\dot{v}_y|\leq0.5$. The reason for specifying left-derivatives for the predicates inside the A and B is to avoid an aggressive entrance into them. The regions in two-dimensional $x-y$ plane can be represented with linear predicates as follows:
\begin{equation}
\ 
    \begin{split}
        &R_A=x\geq 1.5\wedge x\leq 2\wedge y\geq 4.75\wedge y\leq 5.25,\\
        &R_B=x\geq 4\wedge x\leq 5\wedge y\geq 1\wedge y\leq 3,\\
        &R_C=x\geq 2\wedge x\leq 4\wedge y\geq 1\wedge y\leq 5.
    \end{split}
\end{equation}
\label{example_d_int}
\end{example}

\begin{figure}
\centering
\begin{subfigure}{.5\textwidth}
    \centering
    \includegraphics[trim=47 0 50 10,clip, width=\linewidth]{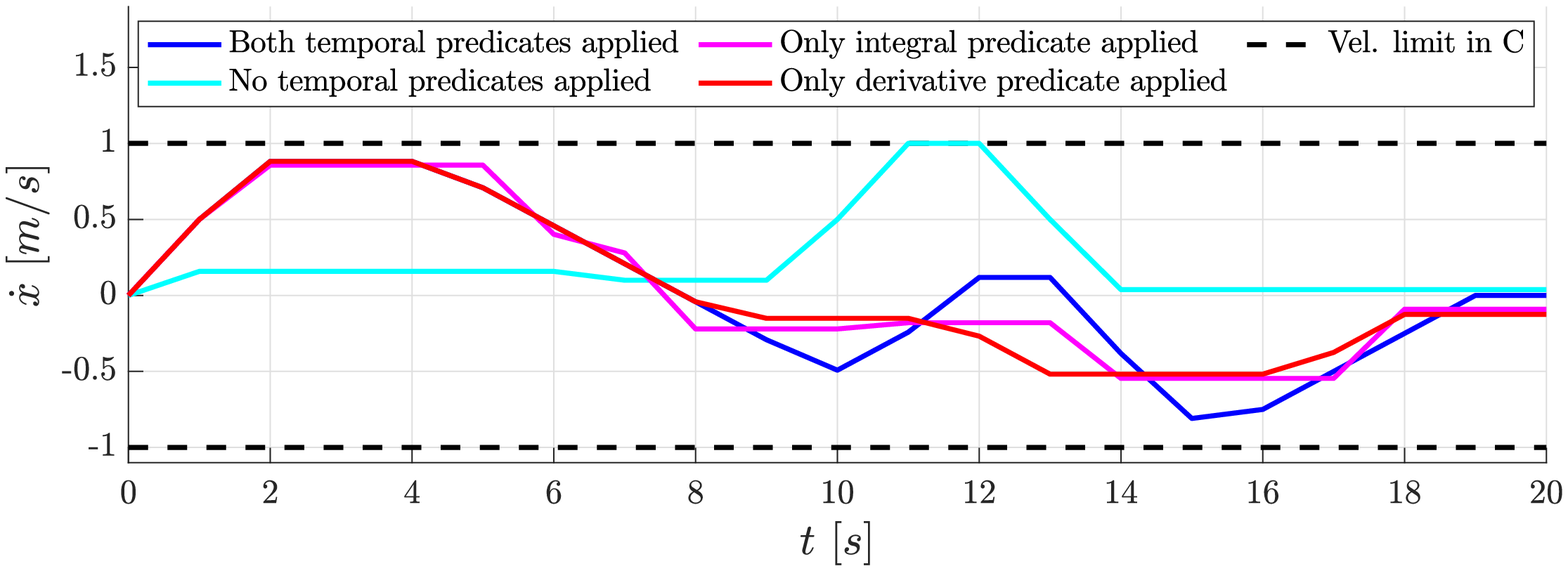}
         \caption{}
         \label{fig:sim_vel_x}
\end{subfigure}
\begin{subfigure}{.5\textwidth}
    \centering
    \includegraphics[trim=47 0 50 10,clip, width=\linewidth]{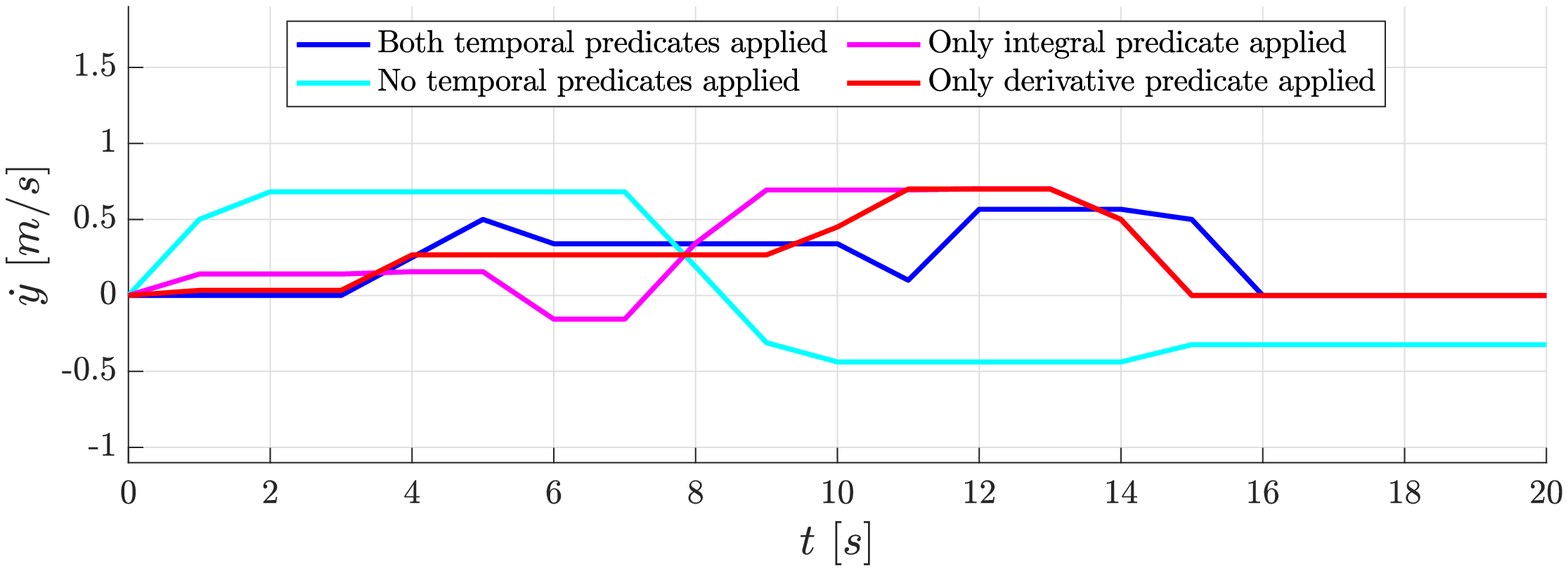}
         \caption{}
         \label{fig:sim_vel_y}   
\end{subfigure}

 \caption{\textcolor{black}{For four different cases, change of velocities with time in a) $x$ and b) $y$ directions. The only trajectory that enters the region C is the cyan one by increasing its speed to $1\ m/s$ in $x$ direction which is a requirement for all cases \eqref{eq:formula}.}}
    \label{fig:sim_vel}
\end{figure}

\begin{figure}[htb!]
   \centering
\begin{subfigure}{.5\textwidth}
    \centering
    \includegraphics[trim=49 0 50 10,clip, width=\linewidth]{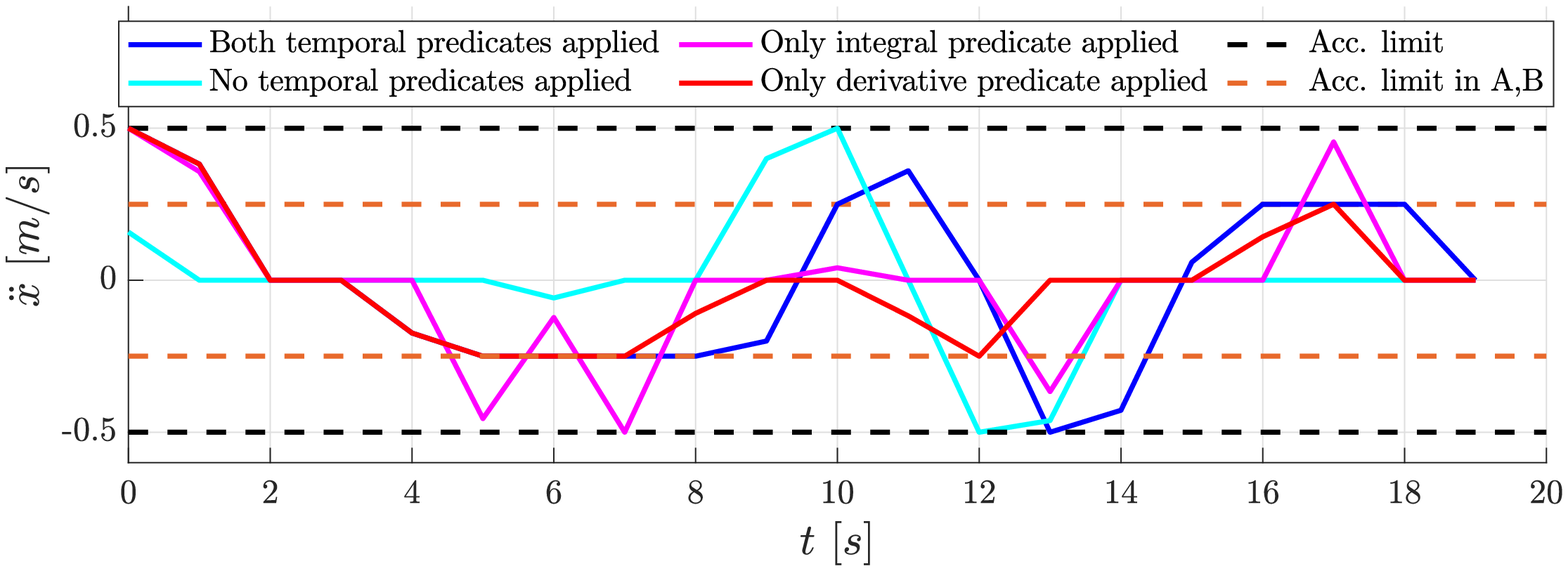}
         \caption{}
         \label{fig:sim_acc_x}
\end{subfigure}
\begin{subfigure}{.5\textwidth}
    \centering
    \includegraphics[trim=49 0 50 10,clip, width=\linewidth]{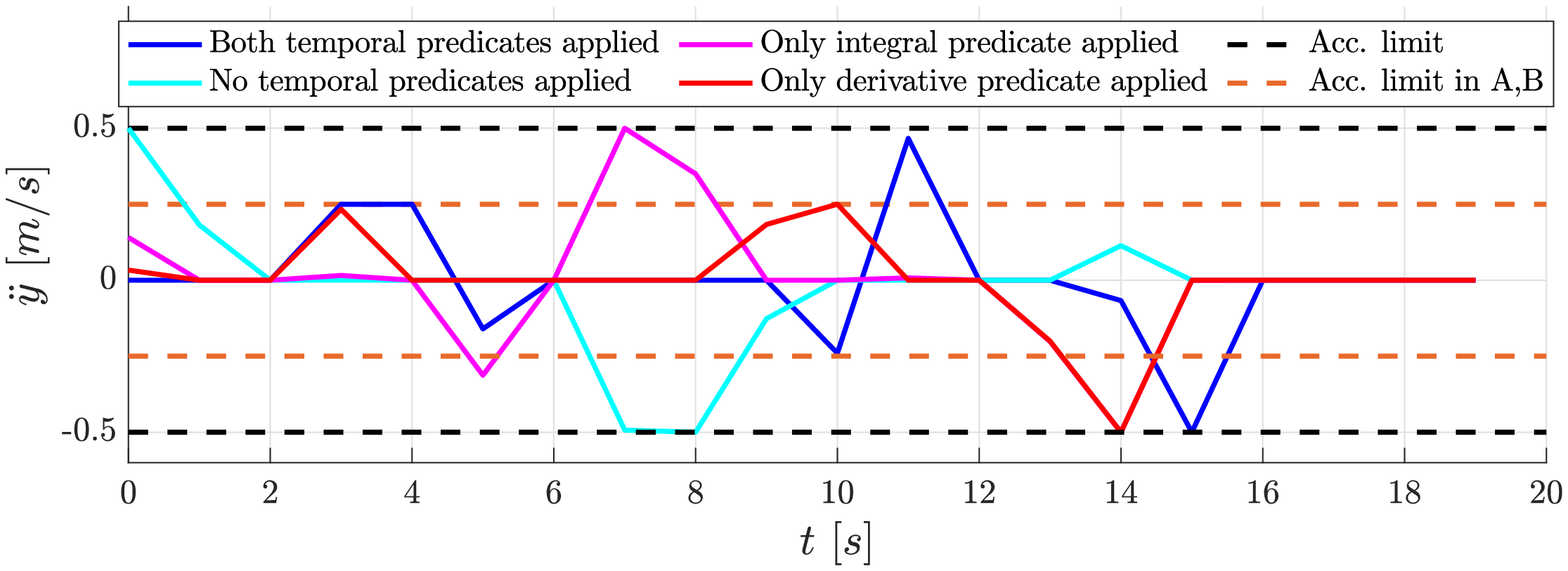}
         \caption{}
         \label{fig:sim_acc_y}   
\end{subfigure}
    \caption{\textcolor{black}{Change of accelerations with time in a) $x$ and b) $y$ directions, calculated via the right-derivative of the velocities. For all the four cases the general limit of $0.5\ m/s^2$ is obeyed. Moreover, blue and red trajectories satisfy a stricter limit of $0.25\ m/s^2$ defined by the left-derivative of the velocities inside the areas A and B.}}
    \label{fig:sim_acc}
\end{figure}

To bound the total traveled distance, we define integral predicates (Def. \eqref{definition_temp_pred}) that will keep the distance in each particular direction at least $2\ m$ by integrating the speeds in that direction. Moreover, we also constrained the maximum acceleration in any direction using derivative predicates defined in Def. \ref{definition_derivative_pred}. Although one can limit the control inputs to do that in this particular example, where the inputs are the acceleration, it is not generally the case for other systems. It is also worth to mention that the expressions with the absolute value operators in \eqref{eq:formula} can be represented by multiple linear inequalities.

\subsection{Simulations}\label{simulations}
\textcolor{black}{We develop a control synthesis tool that generates trajectories satisfying the given temporal logic specifications including the new predicates. Specifications are not necessarily in negation-free form unlike \cite{akazaki2015time}, \cite{rodionova2016temporal}, and \cite{haghighi2019control}, in which the robustness degree is used directly in the algorithms. The global specification is encoded as binary constraints as described in Sec. \ref{solution approach}.} YALMIP \cite{yalmip} is used to model the problem in \eqref{eq:z_optimization}, which is solved via Gurobi \cite{gurobi} in MATLAB R2019b. A laptop computer with 1.8 GHz, Intel Core i5 processor is used to run the simulations with $\delta t=1\ s$..

First, the system trajectory is computed according to the original specification \eqref{eq:formula}. Then, for comparison, the system trajectories are computed to satisfy specifications 1) without the integral and derivative predicates defined in the green areas, 2) without the integral predicates, $\mu^{i,1}_{[0,6]}$ and $\mu^{i,2}_{[0,6]}$, and 3) without the derivative predicates,  $\mu^{d,1}_-$ and $\mu^{d,2}_-$. The generated trajectories are illustrated in Fig.~\ref{fig:simulation}. It is evident from the figures that with the integral predicates, the robot does not only service B but also it is required to travel inside the region more to monitor the natural habitat better. Moreover, the derivative predicates constrain the acceleration of the robot successfully both inside and outside of the regions. For instance, compared to no integral and derivative predicates case, given in cyan in Fig. \ref{fig:sim_none}, when region-based derivative constraint is applied, red in Fig. \ref{fig:sim_deriv}, robot cannot accelerate enough to enter region C and go around instead. \textcolor{black}{This can also be observed in Fig. \ref{fig:sim_vel_x} where the only trajectory that enters the C is the cyan one by increasing its horizontal velocity to $1\ m/s$ inside it. A similar effect of the derivative predicate can be seen in Fig. \ref{fig:sim_acc} in which the trajectories without it (cyan and magenta) violates the acceleration limit ($0.25\ m/s^2$) in each direction inside the regions A and B. On the other hand, all four trajectories obey the general acceleration limit of $0.5\ m/s^2$.} Finally, when there is no derivative predicate applied, the trajectory in magenta also avoids C since it has to slow down inside B to travel the minimum required distance. This makes the magenta trajectory a more aggressive one compared to the blue trajectory with both integral and derivative predicates inside B. The comparison of these four scenarios in terms of the control effort and computation time are also presented in Table \ref{tab:results}.

\renewcommand{\arraystretch}{1.35}
\begin{table}[htb!]
    \centering
        \caption{Results of the simulation scenarios in Fig. \ref{fig:simulation}.}
\begin{tabular}{c||c|c|c|c}
 & \begin{tabular}{@{}c@{}}Int.\&Der.\\ Pred.\\(blue)\end{tabular} & \begin{tabular}{@{}c@{}}Only Int.\\ Predicates\\(magenta)\end{tabular} & \begin{tabular}{@{}c@{}} Only Der.\\ Predicates\\(red)\end{tabular} & \begin{tabular}{@{}c@{}}None of\\Them \\(cyan)\end{tabular}  \\\hline
    \begin{tabular}{@{}c@{}}Total Cost\\ $\mathcal{J}
       ^*$ $[-]$\end{tabular} &$6.5363$ &$4.8253$ &$4.0749$ &$3.9930$\\\hline
     \begin{tabular}{@{}c@{}}Soln.\\ Time $[s]$\end{tabular} &$4.53$ &$15.43$ &$1.13$  &$2.29$\\
    \end{tabular}
    \label{tab:results}
\end{table}
\renewcommand{\arraystretch}{1}

\section{Conclusion and Future Work} \label{conclusion}

In this study, we introduce integral and derivative predicates for Signal Temporal Logic. These new predicates can be used to define specifications regarding the cumulative effects and the rate of change in a signal. We show that the integral and derivative predicates can be encoded as mixed-integer linear constraints. Optimal trajectories satisfying the complex temporal logic specifications are then obtained by solving a mixed-integer linear program. A case study on the control of an autonomous robot in two-dimensional continuous space is included. We show how the new predicates can be used in the design of the mission specifications in a richer and more expressive way compared to the conventional STL. As a future direction, we plan to extend this work to multi-agent systems that are coupled through the specifications of their objectives and constraints (e.g., \cite{Bhat19,Peterson20,buyukkocak2021planning}).

\bibliography{acc21_arxiv_final}

\end{document}